\documentclass[conference]{IEEEtran}

\usepackage{cite}
\ifCLASSINFOpdf
    \usepackage[pdftex]{graphicx}
    \graphicspath{{../Figures/}}%{../pdf/}{../jpeg/}
    \DeclareGraphicsExtensions{.pdf,.jpeg,.png}
\else
\fi
\usepackage{amsmath}
\usepackage{amsfonts}
\usepackage{algorithmic}
\usepackage{array}
\ifCLASSOPTIONcompsoc
  \usepackage[caption=false,font=normalsize,labelfont=sf,textfont=sf]{subfig}
\else
  \usepackage[caption=false,font=footnotesize]{subfig}
\fi

\usepackage{xcolor}

\newtheorem{theorem}{Theorem}[section]

\makeatletter  %for the first page footer
\def\ps@IEEEtitlepagestyle{%
  \def\@oddfoot{\mycopyrightnotice}%
  \def\@evenfoot{}%
}
\def\mycopyrightnotice{%
  {\footnotesize 
  \begin{minipage}{\textwidth}
  \centering
  978-1-5090-3009-5/17/\$31.00 \textcopyright 2017 IEEE  %<--Change
  \end{minipage}
  }
}

\begin{document}

\title{User Activity Detection in Massive Random Access: Compressed Sensing vs. Coded Slotted ALOHA}

\author{\IEEEauthorblockN{Veljko Boljanovi\' c, Dejan Vukobratovi\' c}
\IEEEauthorblockA{Department of Power, Electronic and Communications Engineering\\ 
University of Novi Sad, Serbia\\
Email: veljkoboljanovic@gmail.com, dejanv@uns.ac.rs}
\and
\IEEEauthorblockN{{Petar Popovski, \v Cedomir Stefanovi\' c}
\IEEEauthorblockA{Department of Electronic Systems\\
Aalborg University, Denmark\\
Email: \{petarp,cs\}@es.aau.dk}}}

%Telephone: (800) 555--1212, Fax: (888) 555--1212}

\maketitle

\begin{abstract}
Machine-type communication services in mobile cellular systems are currently evolving with an aim to efficiently address a massive-scale user access to the system.
One of the key problems in this respect is to efficiently identify active users in order to allocate them resources for the subsequent transmissions.
In this paper, we examine two recently suggested approaches for user activity detection: compressed-sensing (CS) and coded slotted ALOHA (CSA), and provide their comparison in terms of performance vs resource utilization.
Our preliminary results show that CS-based approach is able to provide the target user activity detection performance with less overall system resource utilization.
However, this comes at a price of lower energy-efficiency per user, as compared to CSA-based approach.   
\end{abstract}

\IEEEpeerreviewmaketitle

\section{Introduction} \label{intro}
Reservation-based access protocols are standardly used in mobile cellular systems, where the first step is detection/identification of the (a priori unknown) set of active users, based on which the assignment of time-frequency resources and the subsequent data transmissions take place, cf. \cite{3GPPTS36.213,3GPPTS36.321}.
The user activity detection in mobile cellular networks is slotted ALOHA (SA) based, which is a random access algorithm characterized by the simplicity of implementation, while suffering from low performance. Specifically, the throughput of slotted ALOHA in the basic, collision channel model, is upper bounded by $1/e$. Such performance may prove sufficient when the load of the cellular access network (i.e., the number of accessing devices) is low; however, it will pose a significant bottleneck for the foreseen number of accessing devices in the scenarios pertaining to the Internet-of-Things \cite{M2M}.

The main reason behind the modest throughput of slotted ALOHA is due to the resources wasted on collision slots. In particular, it is assumed that the access point is unable to decode any transmission occurring in a collision slot, making use only of the slots containing a single transmission (singleton slots). Obviously, throughput performance could be boosted through development of reception techniques that are able to exploit collisions, complemented by transmission strategies that foster such operation.
%I.e., high throughput in random access scenarios can be achieved with a combination of a properly designed transmission protocol and a multi-user detection technique.

In this paper, we review two advanced approaches for user activity detection in random access scenarios, both being candidate solutions to be used in 5G access networking. The first approach is compressed-sensing (CS) based \cite{CS-MUD}, and the second is coded slotted ALOHA (CSA) \cite{PSLP2014}. 
%Both approaches are used when the number of active users is significantly lower than the total number of users in the cell; however, the manner in which they operate differs significantly. 
In CS, all active users transmit simultaneously during the activity detection phase (i.e., their transmissions collide/overlap completely), where a transmitted signal uniquely identifies the user transmitting it, and the common receiver uses an CS recovery algorithms for the reconstruction of the set of active users. On the other hand, in CSA, users transmit replicas of their unique identifiers in randomly chosen slots of the activity detection phase. The receiver relies on singleton slots to start decoding process, which is then upheld by application of successive interference cancellation (SIC) that removes replicas of decoded transmissions, potentially reducing collision slot to singletons and enabling new round of decoding. We discuss the key assumptions and characteristics of both frameworks and compare their efficiency in terms of resources and energy-expenditure required to achieve target detection performance. Asymptotic performance limits are reviewed and finite-length numerical simulations are provided for both frameworks. Preliminary results show that compressed-sensing based approaches provide the required user activity detection performance with less overall resource utilization. However, this comes at a price of exploiting a user behavior that leads to higher per-user resource utilization, and thus lower energy-efficiency, as compared to coded slotted ALOHA approach.

The rest of the paper is organized as follows. In Sec.~II, we provide background on CS and CSA approaches that underlie the considered user activity detection schemes. In Sec.~III, CS and CSA-based approaches for user activity detection are presented and placed within a common framework. Sec.~IV reviews asymptotic results, while Sec.~V provides numerical results. Finally, Sec.~VI concludes the paper.

\section{Background} \label{background}

\subsection{Compressed Sensing} \label{CS_back}

%The traditional sampling approach is based on \cite[Th. 1]{Shannon}: the sampling rate should be at least twice the maximum frequency present in the signal.

The CS framework allows undersampling the signal and then reconstructing it successfully, given that the signal being sampled is sparse, either in its canonical form, or in some transform domain.
%In other words, CS framework assumes that the sampled signal contains only a fraction of non-zero components \cite{CS_Donoho}.
More precisely, suppose that $\mathbf{x} \in \mathbb{C}^{N}$ is a $k$--sparse signal, i.e., it contains $k$ non-zero components, and $\mathbf{y} \in \mathbb{C}^{m}$ is its sampled version. Then, the ideal (noiseless) CS undersampling process can be expressed in the following way:
\begin{equation} \label{cs_eq}
\mathbf{y}=\mathbf{A}_{cs}\mathbf{x}
\end{equation}
where the sampling matrix $\mathbf{A}_{cs} \in \mathbb{C}^{m \times N}$ models the sampling process.
The design of the matrix $\mathbf{A}_{cs}$ is the fundamental problem of CS.
Specifically, $\mathbf{A}_{cs}$ should have the properties such as coherence or restricted isometry property.
Some simple ensembles of random matrices, e.g., those with real entries sampled from Gaussian or binary entries sampled from Bernoulli distribution, are among suitable choices for $\mathbf{A}_{cs}$. 

The expression (\ref{cs_eq}) represents an underdetermined system of linear equations ($m<N$), solvable due to the sparsity of the vector $\mathbf{x}$. Note that the vector $\mathbf{y}$ is a linear combination of $k$ columns of matrix $\mathbf{A}_{cs}$.
Thus, to reconstruct $\mathbf{x}$ from its samples $\mathbf{y}$, we need to solve the inverse problem: i) to find the support set of $\mathbf{x}$ (i.e., the indices of the $k$ columns included in the linear combination), and ii) the non-zero values of $\mathbf{x}$ (i.e., the weights that define the linear combination).
A basic optimization method for solving (\ref{cs_eq}) is $l_0$-minimization:
\begin{equation} \label{l0_min}
\min \left\{ \left\| \mathbf{z} \right\|_0  :  \mathbf{A}_{cs}\mathbf{z}=\mathbf{y} \right\}
\end{equation}
where $\left\| \mathbf{z} \right\|_0$ is the number of nonzero elements of vector $\mathbf{z}$.
Solving (\ref{l0_min}) is a known NP-hard problem \cite{CS_math}.
Alternatively, the relaxation of $l_0$-minimization is used, called $l_1$-minimization or basis pursuit, described in \cite{BP}. 

In this paper, for CS reconstruction, we use well-known greedy iterative algorithm called Orthogonal Matching Pursuit (OMP) \cite{OMP}.
OMP aims to find indices of nonzero elements of $\mathbf{x}$ by adding one index to the current support set at each iteration.
In addition, proper choice of indices minimizes the residual which eventually leads to the convergence of OMP.
Our choice for OMP is motivated by its well-understood performance, including its asymptotic behavior in terms of sufficient number of samples required for reliable detection \cite{Tropp}. However, one can easily replace OMP in our arguments in the rest of the paper with some other CS recovery methods, such as approximate message passing (AMP) \cite{AMP}.
Finally, we note that the model in (\ref{cs_eq}) can be easily extended to a noisy scenario, where additive Gaussian noise vector $\mathbf{n} \in \mathbb{C}^{m}$, whose components are i.i.d. Gaussian entries with variance $\sigma^2$, is added to the right-hand side of (\ref{cs_eq}). 

\subsection{Coded Slotted ALOHA} \label{CSA_back}

CSA is an umbrella term for SA-based family of random access protocols in which: (i) users contend by transmitting replicas of their packets sent in randomly selected slots, (ii) each packet replica contains pointers to all other replicas of the same packet, (ii) once a replica is decoded, the IC is used to remove the other related replicas, thereby propelling new iterations of decoding and replica removal \cite{PSLP2014,IRSA}.
%Contention resolution extended well known schemes in which each user sends multiple transmission replicas in one frame.
In the basic variant of CSA, called irregular repetition slotted ALOHA (IRSA), the slots are organized in frames, and users contend by selecting a random subset of slots in a frame in which replicas are transmitted.
It was shown that, for the collision channel model, the described iterative successive interference cancellation (SIC) process is analogous to iterative erasure decoding of low-density parity check (LDPC) codes, motivating the use of modern coding theory tools to design and analyze CSA schemes.
In the rest of the paper, we restrict our attention to IRSA.
In the idealized scenario, we assume that a singleton slot is reliably decoded and that the SIC is ideal.

\begin{figure}[!t]
\centering
\includegraphics[width=2.3in]{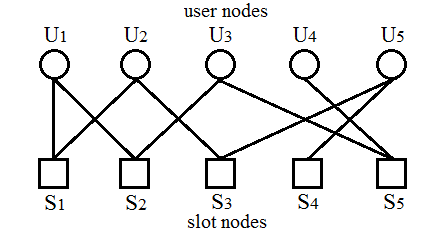}
\caption{An example of the bipartite graph with $k=5$ active users and $M=5$ slots. The process of decoding starts off with detection of the singleton slot S\textsubscript{4}. When replica of user U\textsubscript{5} is successfully decoded, interference from user U\textsubscript{5} can be eliminated from all corresponding slots, i.e. edges from U\textsubscript{5} to S\textsubscript{3} and S\textsubscript{4} can be removed. The next iteration starts similarly -- slot S\textsubscript{3} is detected as the singleton slot and replica of user U\textsubscript{2} is decoded. Consequently, interference from U\textsubscript{2} is eliminated from S\textsubscript{1} and S\textsubscript{3}. The process iterates until all edges are removed or none of singleton slots can be detected.}
\label{bgraph}
\end{figure}

For the design of IRSA scheme, it is useful to represent the contention process via a bipartite graph, see Fig.~\ref{bgraph}, analogous to graph representation of LDPC codes \cite{LDPC_Gal}.
The graph contains user nodes, one for each of the $k$ active users, and slot nodes, one for each of $M$ slots in the frame, and the edges correspond to transmissions of replicas.
The number of replicas is determined by a user degree distribution $\Lambda(x)=\sum_i \Lambda_i x^i$, where $\Lambda_i, 0 \leq i \leq M,$ represents the probability that a user sends $i$ replicas. The average number of replicas per user is denoted by $\overline{\Lambda}$.
The iterative SIC decoder that ``peels off'' the graph represents the process of transmission detection and decoding.
Note also that, in this paper, we will represent the relationship between replicas and slots via adjacency matrix $\mathbf{A}$, where columns correspond to users and rows correspond to slots.
The element $a_{i,j}$ of the matrix $\mathbf{A}$ is equal to one if the $j$-th user sends a replica in the $i$-th slot, otherwise, it is equal to zero.

\section{User Activity Detection: CS vs. CSA} \label{CSvsCSA}

\subsection{System Model and Problem Formulation} \label{sys_model}

As already outlined, the role of the user activity detection is to extract the set of active users in a cell in order to allocate them resources.
We consider a model with a single receiver and $N$ users, out of which $k$ randomly selected users are active. In the user detection phase, we assume active users communicate to the receiver their unique identification codeword $\mathbf{a}_i, 1 \leq i \leq N$, where a user-to-codeword association is known to the receiver. Each active user sends its own codeword simultaneously with other active users within the time frame of duration $T_F=m$ symbols. The time frame period is divided into $M$ slots, each slot containing $T_S=m/M$ symbols. We assume ideal synchronization among the users and the receiver is established at the symbol, slot and frame level. The received signal $\mathbf{y}$ at the receiver can be expressed as:
\begin{equation} \label{model_eq}
\mathbf{y}=\mathbf{A}_t\mathbf{x}+\mathbf{n},
\end{equation}
where $\mathbf{A}_t \in \mathbb{C}^{m \times N}$, $t=\left\{ cs,csa \right\}$, is a matrix defined by the applied framework (explained later in the section), $\mathbf{x} \in \mathbb{C}^N$ is the vector that determines the set of active users, and $\mathbf{n} \in \mathbb{C}^N$ is the additive Gaussian noise. The vector $\mathbf{x}$, which is equivalent in both frameworks, is $k$-sparse and its nonzero elements describe the active users. Thus, finding the support set of vector $\mathbf{x}$ is equivalent to the user activity detection. The nonzero elements may contain information about the channel gains, but for simplicity, we assume existence of the (ideal) uplink power control, which ensures all nonzero elements of $\mathbf{x}$ are equal. Consequently, in the noiseless scenario, vector $\mathbf{y}$ represents a scaled sum of columns of matrix $\mathbf{A}_t$.

In the following, through the definition of $\mathbf{A}_{t}$, we investigate CS-- and CSA--based approaches for user activity detection. Our goal is to minimize resources required for reliable user activity detection, as detailed in Section \ref{asy_res}.

\subsection{CS-based Approach for User Activity Detection} \label{Frame_CS}

In the CS-based user activity detection, introduced in \cite{FRG}, the model presented in (\ref{model_eq}) is specialized as follows. The signal at the receiver is given by (\ref{model_eq}), where $\mathbf{A}_{t}=\mathbf{A}_{cs}=[\mathbf{a}_1 \mathbf{a}_2 \ldots \mathbf{a}_N] \in \mathbb{C}^{m \times N}$ is the matrix containing user identification codewords as columns. The time frame is not slotted, i.e., $M=1$, and every user's codeword $\mathbf{a}_i, 1 \leq i \leq N$ is of length $m$ symbols. For simplicity, we assume the codewords are randomly-generated binary, BPSK modulated sequences of $m$ BPSK symbols $\{+1,-1\}$. Thus $\mathbf{A}_{cs} \in \mathbb{R}^{m \times N}$ is a Bernoulli matrix with entries $\pm 1$. In this paper, for convenience, the columns are normalized to unit norm. In this setup, the model in (\ref{model_eq}) becomes equivalent to the noisy version of the CS model in (\ref{cs_eq}). Thus, using CS methodology, we can reconstruct the signal $\mathbf{x}$ and recover the set of active users (i.e., the support set of $\mathbf{x}$) using the CS recovery algorithms such as OMP. Moreover, one can use theoretical results to provide asymptotic analysis on the length of user codewords $m$ required for reliable signal recovery, as reviewed in Section \ref{asy_res}.

\subsection{CSA-based Approach for User Activity Detection} \label{Frame_CSA}

\begin{figure}
\centering
\includegraphics[width=2.3in]{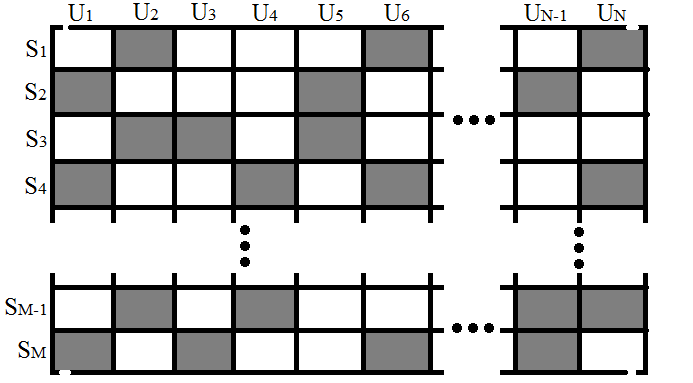}
\caption{An example of specific structure of matrix $A_{csa}$. Replicas of users' ID sequences are represented with grey color. Columns correspond to users and rows correspond to slots.}
\label{Acsa}
\end{figure}

In the CSA-based user activity detection, the model presented in (\ref{model_eq}) assumes different format of the user identification codewords, and consequently, different form of the matrix $\mathbf{A}_{t}=\mathbf{A}_{csa}$. In order to define the format of user identification codewords $\mathbf{a}_i$'s, we introduce the division of the time frame into $M$ slots, where the length of each slot equals $T_S=\lceil \log_2(N) \rceil$ (binary) symbols (for convenience, we assume $T_S$ divides $m$). In addition, each user is uniquely identified by its ID sequence $\mathbf{s}_i$ of length $l_{s}=\lceil \log_2(N) \rceil$ binary symbols that exactly fits a single time slot. To generate $\mathbf{a}_i$, the $i$-th user applies CSA approach, namely, it first generates the degree $d_i$ from $\Lambda(x)$, randomly samples $d_i$ out of $M$ available slots in the time frame, and transmits $d_i$ replicas of its own ID sequence $\mathbf{s}_i$ within the selected slots. Thus in general, a user identification codeword $\mathbf{a}_i$ consists of length-$T_S$ sub-blocks, where only $d_i$ sub-blocks are BPSK modulated symbol sequences, while the remaining $M - d_i$ represent all-zero sequences. Arranging all the user identification codewords as the columns of the matrix $\mathbf{A}_{csa}$, one obtains the structured matrix of the form illustrated in Fig. \ref{Acsa}. Finally, upon receiving the frame, the receiver decodes the user transmissions using iterative SIC decoder and eventually recovers user ID sequences from which the set of active users is reconstructed. 

\subsection{A Remark on User Activity Detection}

We conclude this section with a general remark that detection of the active users can be made by exploitation both of the observations of the activity patterns and the received information.
However, in CS, all active users are transmitting throughout the detection phase, i.e., their activity patterns are the same and thus do not carry any specific information.
On the other hand, in CSA, the activity pattern of a user can be related to its identity, e.g., its identity can be used as the seed of a random number generator used to select slots in which the user is active.
Yet, this prospect is typically not exploited, and the user identity is communicated only through information sent in the packet replicas; this approach is used in the paper.
% We continue by introducing a framework that can embrace both CS- and CSA-based approaches, which is then instantiated for each of them separately.     

\section{Asymptotic Results} \label{asy_res}

In this section, we compare the two frameworks in terms of required system resources needed to achieve asymptotically vanishing probability of error of user activity detection. The required resources are effectively equal to the length (i.e., the number of symbols) of user identification codewords. However, in the CSA framework, from a single user perspective, the majority of the $M$ slots contain no signal, motivating us to consider energy efficiency as a comparison metric, by taking into account number of the actual symbol transmissions. We start by a review of the known asymptotic bounds on the codeword lengths required for the reliable detection.

\subsection{Noiseless Model} \label{noiseless_asy}

\textbf{CS-based framework:} As mentioned in Section \ref{background}, sampling matrices with i.i.d. Gaussian or Bernoulli entries as suitable choice for the CS framework.
Most of results for OMP are based on certain assumptions about coherence among columns of matrix $\mathbf{A}_{cs}$. The coherence threshold, which ensures exact reconstruction of vector $\mathbf{x}$, was evaluated in \cite{TroppGreed}.
However, as noted in \cite{Tropp}, OMP would require an enormous number of measurements (symbols) in that case. A useful performance guarantee for OMP follows from the result below \cite[Th. 6]{Tropp}.
\begin{theorem} \label{cs_th}
Fix $\delta \in \left(0,0.36 \right)$, and choose $m \geq Ck \ln\left(N/ \delta \right)$ where $C$ is a constant. Suppose that $\mathbf{x}$ is an arbitrary $k$-sparse signal in $\mathbb{R}^N$, and draw a random $m \times N$ Gaussian/Bernoulli\footnote{Note that the statement of the theorem holds for a more general class of so called admissible matrices defined in \cite{Tropp}.} measurement matrix $\mathbf{A}_{cs}$ independent from the signal. Given the data $\mathbf{y}=\mathbf{A}_{cs}\mathbf{x}$, OMP can reconstruct the signal with probability exceeding $1- \delta$.
\end{theorem}

The above result has been further refined in \cite{GaussianCase} for the case of Gaussian matrices, suggesting that $C \leq 20$. In asymptotic setting, i.e., when $N, k \rightarrow \infty$, it is possible to reduce the constant to $C \leq 4+ \eta$, where $\eta$ is a positive number. Thus, the probability of imprecise support set estimation is reduced below $\delta$ if the number of measurements satisfies:
\begin{equation} \label{CS_asymp}
m \geq 4k \ln \left(N/ \delta \right).
\end{equation}
Numerical simulations in \cite{Tropp} revealed that (\ref{CS_asymp}) is slightly pessimistic and it turned out that OMP requires fewer measurements. For example, if $\delta=0.01$ in the models with $N=256, 1024$ users, required number of measurements is approximately $m \approx 2k \ln \left(N \right)$, instead of $m = 4k \left( \ln \left(N \right)+4.6 \right)$. Admittedly, as $\delta$ decreases, OMP requires more measurements, i.e. the expression for $m$ changes. Furthermore, numerical evidence point out that constants for the Gaussian matrices  are closely matched by the case of Bernoulli matrices \cite{Tropp}.

\textbf{CSA-based framework:} In the CSA framework, the average number of active user transmissions per slot, also called the average system load, can be defined as $G=k/M$.

\begin{figure*}[!t]
	\centering
	\subfloat[]{
	\includegraphics[width=0.99\columnwidth]{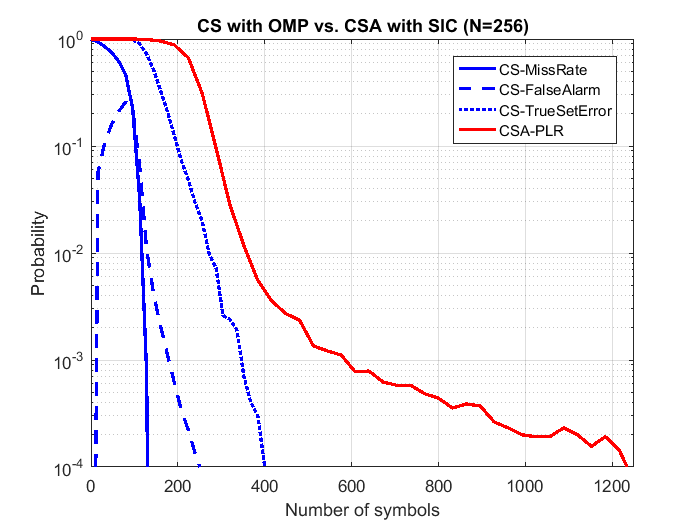}
	\label{256users}
	}
	\hfil
	\subfloat[]{
	\includegraphics[width=0.99\columnwidth]{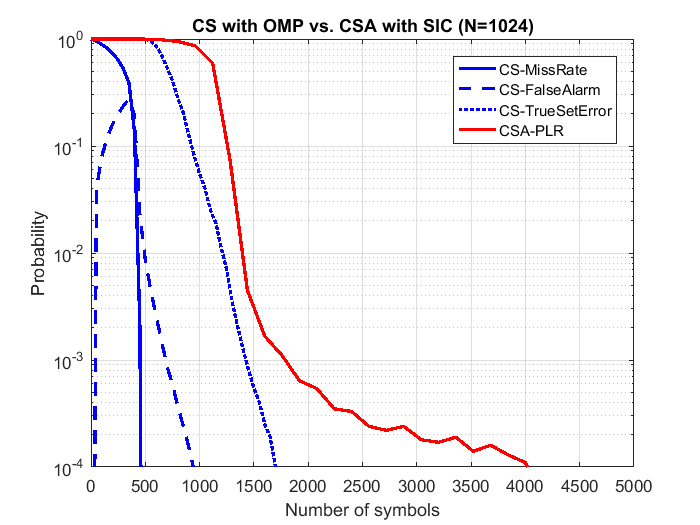}
	\label{1024users}
	}
	\caption{Performance comparison in the noiseless models with $N=256$, $k=25$ (a) and $N=1024$, $k=100$ (b), and the user node degree distribution used in the CSA framework is $\Lambda(x)=0.25x^2+0.6x^3+0.15x^8$.}
\end{figure*}

\begin{figure}
	\centering
	\includegraphics[width=0.99\columnwidth]{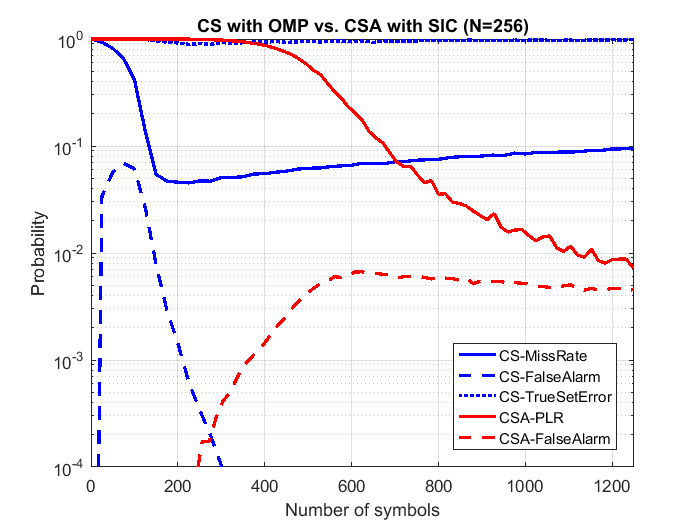}
	\caption{Performance comparison in the noisy model with $N=256$, $k=25$, $\text{SNR}=10~\text{dB}$, and the user node degree distribution used in the CSA framework is $\Lambda(x)=0.25x^2+0.6x^3+0.15x^8$.}
	\label{256users-noise}
\end{figure}

Following \cite{IRSA}, for a fixed $\Lambda(x)$, an asymptotic ($N, m \rightarrow \infty$) threshold $G^*$ is defined such that, if $G \leq G^*$, all transmissions will be successfully detected and decoded with probability approaching one. Otherwise, the probability of successful detection will asymptotically vanish. Therefore, all active users are reliably detected when $\frac{k}{M} \leq G^*$, i.e. when:
\begin{equation} \label{M_greater}
M \geq \frac{k}{G^*}.
\end{equation}
By multiplying both sides with the number of symbols in a single ID sequence $l_{s}$, the total number of symbols $m$ is obtained:
\begin{equation} \label{CSA_asymp}
m = Ml_{s} \geq \frac{k}{G^*} l_{s}.
\end{equation}
For example, for the node-oriented distribution $\Lambda(x)=0.25x^2+0.6x^3+0.15x^8$, the threshold equals $G^*=0.892$ and thus, (\ref{CSA_asymp}) becomes $m \geq 1.121k l_{s}$.

\subsection{Noisy Model} \label{noisy_asy}

\textbf{CS-based framework:} Unlike the noiseless model where OMP algorithm iterates until residual $\mathbf{r}$ becomes zero, a different stopping rule for OMP has to be used in the noisy model. Furthermore, the coherence-based results for OMP require additional conditions for the nonzero elements of vector $\mathbf{x}$. A following theorem provides a performance guarantee for the correct detection of the support set of $\mathbf{x}$ in the presence of Gaussian noise \cite[Th. 7]{OMP_noise}.

\begin{theorem} \label{cs_noise_th}
Suppose $\mathbf{n} \sim \mathcal{N}(0, \sigma^2  \mathbf{I_n})$ and $\mu < \frac{1}{2k-1}$ and all the nonzero coefficients of $\mathbf{x}$ satisfy
\begin{equation}
|x_i| \geq \frac{2 \sigma \sqrt{m + \sqrt{m \log(m)}}}{1-(2k-1) \mu}.
\end{equation}
Then OMP algorithm which utilizes the stopping rule $\left\| \mathbf{r} \right\|_2 \leq \sigma \sqrt{m + \sqrt{m \log(m)}}$ selects the true support set of $\mathbf{x}$ with probability at least $1-1/m$.
\end{theorem}

%Unlike the noiseless model where OMP algorithm iterates until residual $\mathbf{r}$ becomes zero, a different stopping rule for OMP has to be used in the noisy model. Furthermore, the coherence-based results for OMP require additional conditions for the nonzero elements of vector $\mathbf{x}$. A following theorem provides a performance guarantee for the correct detection of the support set of $\mathbf{x}$ in the presence of $l_2$-bounded noise $\left( \left\| \mathbf{n} \right\|_2 \leq b_2 \right)$ \cite[Th. 1]{OMP_noise}.}

%\VB{\emph{Theorem:} Suppose $\left\| \mathbf{n} \right\|_2 \leq b_2$ and $\mu < \frac{1}{2k-1}$. Then OMP with the stopping rule $\left\| \mathbf{r} \right\|_2 \leq b_2$ recovers exactly the true support set of vector $\mathbf{x}$ if all the nonzero coefficients $x_i$ satisfy $|x_i| \geq \frac{2b_2}{1-(2k-1) \mu}$.}\\

\textbf{CSA-based framework:} In general, the CSA framework implicitly assumes that the noise is addressed by (ideal) physical layer forward error correction (FEC) \cite{IRSA}. However, for the sake of counting $m$, we need to include the FEC rate penalty in our model. Asymptotically, as $N,k \rightarrow \infty$, the rate penalty follows directly from Shannon's channel coding theorem. Thus, in the case of binary-input additive white Gaussian noise (BI-AWGN) channel considered here, the additional multiplicative constant $1/C(\sigma)$ needs to be added to equation \eqref{CSA_asymp}, where $C(\sigma)$ is the capacity of BI-AWGN channel with parameter $\sigma$, to account for the physical-layer FEC. For the non-asymptotic scenario, one can apply bounds following recent advents in finite-length information theory, however, we leave that case for future work.   

\section{Numerical Simulation} \label{num_sim}

The goal of the receiver is to precisely determine the set of active users and thus correctly estimate the support set of the vector $\mathbf{x}$. For the finite number of measurements in the CS framework, the performance of the receiver is analytically intractable and is usually obtained using numerical simulations. Similarly, the performance in the CSA framework deteriorates if the number of slots is finite, with respect to the asymptotic thresholds indicated above.

In the case of limited resources, the reconstruction under OMP algorithm may experience two error events: i) the active user is not detected, and ii) the inactive user is detected as active. Probabilities of these two events are called \emph{missed detection rate} and \emph{false alarm probability}, respectively. On the other hand, in the noiseless CSA framework, there is just a single type of error event -- the active user not being detected. This occurs when none of the user's replicas is successfully detected and decoded, i.e., when the user packet is lost, and the related probability is called \emph{packet loss rate} (PLR). The terms PLR and missed detection rate can be used interchangeably, as they refer to the same type of error event.

Missed detection rate and false alarm probability were considered jointly in \cite{Tropp} as probability of unsuccessful detection of the exact support set of vector $x$. If we were interested in finding the exact support set, we would use that probability in the CS framework, and equivalently, the frame loss rate (FLR) in the CSA framework. However, in access networking, the user activity detection is followed by assignment of resources to the detected users, and the false alarms actually imply resources' loss. Thus, in the paper we separate among false alarms and miss-detections. %\VB{ Treba reci za dodelu resursa i da je bolje posmatrati odvojeno miss rate (PLR) i false alarm}

The numerical analysis is conducted in the models with $N \in \{256, 1024\}$ users, where the number of active users $k$ is 25 and 100, respectively. In the CSA framework, the user degree distribution is set to $\Lambda(x) =0.25x^2+0.6x^3+0.15x^8$ \cite{IRSA}. For both frameworks, the main metric we consider is the total number of symbols $m$ in the transmitted frame necessary for the system to reach sufficiently low missed detection rate.    

The comparison of two frameworks in the noiseless model when $N=256$ and $k=25$ is shown in Fig.~\ref{256users}, and when $N=1024$ and $k=100$ in Fig.~\ref{1024users}; for each case, the results were derived using $10000$ independent Monte Carlo trials. The results show that the CS framework with OMP algorithm is more efficient, i.e., it achieves a given miss detection rate (PLR) with fewer number of symbols $m$. Since miss detection rate quickly tends to zero, the performance of OMP algorithm becomes solely affected by false alarm probability after certain number of symbols. Consequently, probability that the exact support set of vector $x$ is not detected scales similarly to false alarm probability, as presented in Fig.~\ref{256users} and Fig.~\ref{1024users}; it is easy to verify that, when false alarm probability is multiplied with $N-k$, i.e. with the number of inactive users, it becomes practically equal to probability of not finding the exact support set.
On the other hand, PLR in CSA framework starts to decrease rapidly when the number of transmitted symbols $m$ becomes roughly equal to $kl_{s}$, (i.e., when the number of slots $M$ becomes approximately equal to number of the active users $k$), but then, for lower PLRs, it experiences an error floor. Obviously, the length of users' ID sequences in CSA plays a key role; as it is unknown which users will be active, all users are assigned a unique ID sequence of length $l_{s} = \lceil \log_2(N) \rceil$. However, when the probability that a user becomes active is small, using ID sequences with $l_{s} = \lceil \log_2(N) \rceil$ symbols does not seem reasonable, and one can use methods where ID sequence length scales with $k$ instead of $N$. However, in this case, two or more active users may happen to use a same ID sequence, which necessitates a collision resolution in the space of ID sequences and which is out of the paper scope.

The same system model with $N=256$ and $k=25$ is examined in both frameworks in the presence of Gaussian noise $\mathbf{n} \sim \mathcal{N}(0,\sigma^2  \mathbf{I_n})$ and the results are presented in Fig.~\ref{256users-noise}. The results for $N=1024$ and $k=100$ are similar and they are omitted for brevity. In order to fairly compare two frameworks, we define signal-to-noise ratio as SNR per information bit. Since $N=256$ users can be uniquely described with ID sequences $\mathbf{s}_i$ of length $l_{s}=8~\text{bits}$, we can obtain mentioned SNR by dividing SNR per transmitted symbol with code rate in the corresponding framework. The code rates are $R_{cs}= \frac{l_s}{m}$ and $R_{csa}= \frac{1}{\overline{\Lambda}} R_{fec}$ in the CS and the CSA frameworks respectively. The rate $R_{fec}$ is defined as $R_{fec}= \frac{l_s}{c_s}$ which gives $R_{csa}= \frac{1}{ \overline{\Lambda}} \frac{l_s}{c_s}$. The value $c_s$ is the length of ID sequence after FEC coding in the CSA framework (we use Reed-Muller codes with $R_{fec}=1/2$).

In our simulation, we fix $\text{SNR}=10~\text{dB}$.  As Th.~\ref{cs_noise_th} suggests, the absolute value of the nonzero elements of vector $\mathbf{x}$ (i.e. SNR) needs to be sufficiently high in order to allow reliable user activity detection in the CS, even when the number of measurements (symbols) $m$ goes to infinity.
Nevertheless, we present the simulation results for $\text{SNR}=10~\text{dB}$ in order to show that insufficiently high $\text{SNR}$ causes performance deterioration with respect to the noiseless case. Unlike in the noiseless model, probability of erroneous detection of the exact support is dictated here by high miss detection rate. Indeed, low $\text{SNR}$ and a high threshold for the stopping rule from Th.~\ref{cs_noise_th} prevent some active users from being detected. Moreover, when SNR is fixed, increasing code rate in the CS results in poor performance of OMP algorithm.
On the other hand, the CSA code rate remains fixed as $m$ increases and its performance resembles that in the noiseless case. However, noise causes inactive users to be detected as active. Furthermore, the PLR curve is shifted to the right due to the FEC coding exploited in the CSA framework.

It may appear that using the total number of transmitted symbols $m$ as a performance metric to compare CS- and CSA-based framework is, to some extent, unfair. Indeed, in the CS framework, matrix $\mathbf{A}_{cs}$ is a full matrix which means that $k$ active users send a total of $km$ symbols, or $m$ symbols per active user. Therefore, there is a linear relationship between the total number of sent symbols and length of the transmission time frame. In contrast, matrix $\mathbf{A}_{csa}$ from the CSA framework is sparse. E.g., for the used distribution $\Lambda(x)=0.25x^2+0.6x^3+0.15x^8$, the average number of replicas per user is $ \overline{\Lambda}=3.5$, thus, the total number of sent bits is $3.5k \lceil \log_2(N) \rceil$, or $3.5 \lceil \log_2(N) \rceil$ per active user. Clearly, the total number of symbols increases logarithmically with $N$ (and also with $k$ if we assume that $N$ and $k$ scale linearly). Thus, for an arbitrary but fixed $k$, as the number of symbols $m$ grows, the total energy expenditure as well as energy expenditure per active user remains constant in the CSA framework, while it grows linearly in the CS framework. This suggests that for low-delay applications, CS-based framework represents superior choice due to considerably smaller number of symbols $m$ required to achieve satisfactory detection rate (PLR) performance. On the other hand, in the case when the energy resources of the reporting devices are limited, significantly reduced energy consumption per user for CSA-based approach may prove advantageous. 

\section{Discussion and Conclusions} \label{conclusion}

In this paper we examined CS- and CSA-based frameworks for the user activity detection in mobile cellular networks.
We showed that CS-based framework is advantageous in terms of the number of system resources used, implying lower delay of the user activity detection phase.
On the other hand, CSA-based framework involves lower number of transmissions during the activity detection phase on the user basis.

Further extensions may include modification of CSA-based framework, such that the activity pattern and the transmitted information jointly determine the user identity. In this case, the length of the slots in CSA can be reduced, which may also reduce the overall length of the activity detection phase.
On the other hand, this approach would require a reception algorithm that involves operations more complex than SIC.

\section*{Acknowledgment} \label{acknow}

The research presented in this paper was supported in part by the Danish Council for Independent Research, grant no. DFF-4005-00281, and in part by the EU H2020 research and innovation programme under grant agreement no. 734331.

\end{document}